\documentstyle[11pt,aasms]{article}
\slugcomment{Accepted by the Astrophysical Journal}

\begin{document}

\font\twelvei = cmmi10 scaled\magstep1
       \font\teni = cmmi10 \font\seveni = cmmi7
\font\mbf = cmmib10 scaled\magstep1
       \font\mbfs = cmmib10 \font\mbfss = cmmib10 scaled 833
\font\msybf = cmbsy10 scaled\magstep1
       \font\msybfs = cmbsy10 \font\msybfss = cmbsy10 scaled 833
\textfont1 = \twelvei
       \scriptfont1 = \twelvei \scriptscriptfont1 = \teni
       \def\mit{\fam1 }
\textfont9 = \mbf
       \scriptfont9 = \mbfs \scriptscriptfont9 = \mbfss
       \def\bmit{\fam9 }
\textfont10 = \msybf
       \scriptfont10 = \msybfs \scriptscriptfont10 = \msybfss
       \def\bmsy{\fam10 }

\def\etal{{\it et al.~}}
\def\eg{{\it e.g.}}
\def\ie{{\it i.e.}}
\def\lsim{\raise0.3ex\hbox{$<$}\kern-0.75em{\lower0.65ex\hbox{$\sim$}}}
\def\gsim{\raise0.3ex\hbox{$>$}\kern-0.75em{\lower0.65ex\hbox{$\sim$}}}

\title{The MHD Kelvin-Helmholtz Instability:\\
A Two-Dimensional Numerical Study}

\author{Adam Frank \altaffilmark{1}, T. W. Jones \altaffilmark{2},
        Dongsu Ryu \altaffilmark{3}}
\and
\author{Joseph B. Gaalaas \altaffilmark{4}}

\altaffiltext{1}{ Department of Astronomy, University of Minnesota,
    Minneapolis, MN 55455; e-mail: afrank@astro.spa.umn.edu}
\altaffiltext{2}{ Department of Astronomy, University of Minnesota,
    Minneapolis, MN 55455; e-mail: twj@astro.spa.umn.edu}
\altaffiltext{3}{Department of Astronomy and Space Science,
    Chungnam National University,
    Daejeon 305-764, Korea; e-mail: ryu@sirius.chungnam.ac.kr}
\altaffiltext{4}{ Minnesota Supercomputer Institute, University of Minnesota,
    Minneapolis, MN 55455; e-mail: gaalaas@msi.umn.edu}


\begin{abstract}
We have carried out two-dimensional simulations of the nonlinear
evolution of unstable sheared magnetohydrodynamic flows.  These
calculations extend the earlier work of \cite{miura84} and consider
periodic sections of flows containing aligned magnetic fields. Two
equal density, compressible fluids are separated by a shear layer with
a hyperbolic tangent velocity profile. We considered two cases: a
strong magnetic field (Alfv\'en Mach number, $M_a = 2.5$) and a weak
field ($M_a = 5$). Each flow rapidly evolves until it reaches a nearly
steady condition, which is fundamentally different from the analogous
gasdynamic state.  Both MHD flows relax to a stable, laminar flow on
timescales less than or of the order of 15 linear growth times, measured from
saturation of the instability. That timescale is several orders of
magnitude less than the nominal dissipation time for these simulated
flows, so this condition represents an quasi-steady relaxed
state analogous to the long-lived single vortex,
known as ``Kelvin's Cat's Eye'',
formed in two-dimensional nearly ideal gasdynamic simulations of a vortex
sheet.

The strong magnetic field case reaches saturation as magnetic tension
in the displaced flow boundary becomes sufficient to stabilize it.
That flow then relaxes in a straightforward way to the steady, laminar
flow condition. The weak magnetic field case, on the other hand,
begins development of the vortex expected for gasdynamics, but that
vortex is destroyed by magnetic stresses that locally become strong.
Magnetic topologies lead to reconnection and dynamical alignment
between magnetic and velocity fields. Together these processes produce
a sequence of intermittent vortices and subsequent relaxation to a
nearly laminar flow condition in which the magnetic cross helicity is
nearly maximized.  Remaining irregularities show several interesting
properties. A pair of magnetic flux tubes are formed that straddle the
boundary between the oppositely moving fluids. Velocity and magnetic
fluctuations within those features are closely aligned,
representing Alfv\'en waves propagating locally downstream. The flux
tubes surround a low density channel of hot gas that contains most of
the excess entropy generated through the relaxation process.

\end{abstract}

\keywords{hydrodynamics--- magnetohydrodynamics--- turbulence}

\section{Introduction}
Sheared boundary flows are ubiquitous in astrophysical environments as
diverse as the earth's magnetopause and supersonic jets. The
susceptibility of such boundaries to the Kelvin-Helmholtz (KH)
instability is well- known (e.g., \cite{chan61,birk91}).  Development of the KH
instability leads to momentum and energy transport, dissipation,
mixing of fluids and, when magnetic fields are involved, the
possibility of field amplification, reconnection and dynamical
alignment, or self-organization processes.  Although the nonlinear
evolution of the KH instability has been studied in considerable
detail for non-magnetized fluids through extensive analytical calculations
and numerical
simulations (e.g., \cite{corsher84} and references therein),
much less progress has been made in understanding the nature of the KH
instability in magnetohydrodynamics (MHD). That is both because the
MHD KH instability is substantially more complex and because until
recently adequate numerical MHD simulations were difficult.  Yet it is
clear that understanding the physics of the MHD KH instability would
be very useful in a wide variety of contexts.

The basic MHD KH linear stability analysis was carried out a long time
ago (e.g., \cite{chan61}), and it has since been applied to numerous
astrophysical situations including those requiring cylindrical
geometries (e.g., \cite{ferr81,fied84}).
It is clear from these studies that the influence
a magnetic field has on the stability of an initial flow equilibrium
depends on both the strength and orientation of the field.  Obviously,
an equilibrium exists only when the field initially is parallel to the
shear layer.  When the field is aligned also with the flow vectors,
magnetic tension directly inhibits the instability, while if the field is
perpendicular, its role in ideal MHD comes through pressure
contributions modifying the characteristic magnetosonic wave
speed. For equal density, incompressible, inviscid fluid layers
separated discontinuously in a ``vortex sheet'' the linear growth rate
is, $\Gamma = {1\over 2} |{\bmit k} \cdot {\bmit U_o}|[ 1~-~(2 c_a
\hat{\bmit k\cdot} \hat{\bmit B_o})^2/(\hat{\bmit k} \cdot{\bmit
U_o})^2]^{1/2}$, where $\bmit U_o$ is the velocity difference between
the two layers, $c_a$ is the Alfv\'en speed, ${\bmit k}$ is the
wave vector and $\hat{\bmit B_o}$ is the direction of the magnetic
field (\cite{chan61}).  From this, one also can see that a necessary
condition for instability is $c_a < |(\hat {\bmit k} \cdot {\bmit
U_o})/(2 \hat {\bmit k} \cdot \hat{\bmit B_o})|$.
The instability constraint  on the
Alfv\'en Mach
number, $M_a$, is $M_a = U_o/c_a > 2$, when magnetic field, velocity
and perturbation wave vectors line up.  If the shear layer is
of finite thickness, $\Gamma$ will generally fall below the value
predicted above, with a maximum growth rate for wave numbers $k \sim
{1\over a}$, where $a$ measures the width of the transition layer.
Similarly, the finite width of the shear layer stabilizes very short
wavelength modes,${1\over k} << a$, preventing the divergence in the
linear growth rate that
occurs for a discontinuity. It appears that
the same constraint on $c_a$ applies in the finite shear layer (\cite{miupr82},
hereafter ``MP'').  Compressibility of the fluid also tends to reduce
the growth of the instability.  There has been much discussion in the
literature about the existence or nonexistence of an upper limit of
order unity to the sonic Mach number for KH unstable flows (e.g.,
\cite{lan44,miles57,blum70,blumet75,artma87,miura90}).  Especially when
one considers superAlfv\'enic shear layers of finite thickness, it
appears that there is no such limit, although the growth of the
instability under highly supersonic conditions is generally much
slower than the classical, incompressible result.

Even though these results tell us when to expect the MHD KH
instability and when it might have time to develop fully, only
nonlinear calculations can determine its consequences. In recent years
a number of two-dimensional numerical studies have been published
aiming at various aspects of the problem
(e.g.,\cite{tajleb80,miura84,miura87,miura92,wangro84,wu86,belham88,belmon89}),
especially as it applies to the earth's magnetosphere.  However, the
nonlinear physics of MHD KH unstable flows is very complex, and
studies have generally been too limited by computational constraints
(low numerical resolution and insufficient dynamical time coverage) to
answer basic questions definitively about either the important
physical processes involved or the eventual fluid state.

In the present paper we take some steps to help remedy that situation.
Our intent is to address some basic issues identifying the character
of fluid and magnetic field interactions once the MHD KH instability
becomes nonlinear. Thus, we consider only the idealized, periodic KH
instability, postponing for now a study of the evolution of KH
unstable MHD flows in more general and realistic settings.  In this
way, we hope subsequently to understand better how those less
controlled flows are formed. As our starting point we use the very
nice work by \cite{miura84}, who examined the evolution of some normal
mode perturbations to an unstable MHD shear layer.  That problem is
both numerically well posed and leads to a straightforward evolution.
The role of the magnetic field is most critical when it is aligned
with the flow, so we shall focus on that configuration.  This means
that our simulations are strictly two-dimensional, leading to some
important properties in the flows, especially as they apply to the
evolution of the magnetic field.  For example, for planar
symmetry a regenerative dynamo
action cannot develop to amplify the field ({\it e.g.,} \cite{zel57,moff78}).
On the other hand, both 2D and 3D MHD turbulent flows cascade energy to
larger wavenumbers ({\it e.g.,} \cite{pouq78}), whereas 2D hydrodynamical
turbulence produces an inverted energy cascade.
A 2D simulation, however, enables us to achieve
high numerical resolution that can be crucial to obtaining
solutions in which the large scale motions and global statistical
properties are not strongly influenced by numerical diffusion.

Since the character of the flow evolution is rather
different for flows in which the initial magnetic field strength is
large ($c_a \sim U_o$) than when it is small ($c_a << U_o$) we
consider both of those cases.  Miura's calculations were terminated
shortly after the nonlinear saturation of the instability, because of
concerns about numerical limitations and because the periodic version
of the instability is idealized as already mentioned
(cf. \cite{wu86}).  On the other hand, understanding how the flow
relaxes after saturation could be very useful in deciphering the fully
nonlinear behavior of the instability in more general settings.  So,
we have extended our simulations until an approximate equilibrium is
achieved.  We note here that stable, equilibrium flows develop in
these simulations on a much shorter timescale than that required to
dissipate the kinetic energy in the initial configurations.  Miura's
calculations were carried out on a rather coarse numerical grid
($100\times 100$ zones), and he made no effort to examine the
convergence of his results.  On the other hand, the numerical
evolution of complex flows is often highly dependent on numerical
diffusion that effectively limits the Reynolds numbers in the flow.
\cite{maclet94} recently emphasized the special importance of
this concern for MHD flows.  Thus, we have carried out each of our
simulations at two or more resolutions to provide some information
about that issue. We note that \cite{maletal95} have independently
concluded a similar numerical study using different methods. Their
results are entirely consistent with ours and their discussion
complements this one.

The plan of the paper is as follows. In \S 2 we describe our numerical
methods and the initial conditions used.  Section 3 contains a
discussion of the characteristics of the several phases of the
evolution in the instability, while \S 4 is a brief summary with
conclusions.

\section{Methods and Initial Conditions}

MHD describes the behavior of the combined system of a conducting
fluid and magnetic field in the limit that the displacement current
and the separation between ions and electrons are neglected.  So, the
MHD equations represent coupling of the equations of fluid dynamics
with the Maxwell's equations of electrodynamics.  When the effects of
viscosity, electrical resistivity and thermal conductivity can be
neglected on large scales one works with the ideal, compressible MHD equations,
\begin{equation}
{\partial\rho\over\partial t} + {\bmsy\nabla}\cdot
\left(\rho {\bmit u}\right) = 0,
\label{masscon}
\end{equation}
\begin{equation}
{\partial{\bmit u}\over\partial t} + {\bmit u}\cdot{\bmsy\nabla}
{\bmit u} +{1\over\rho}{\bmsy \nabla}p - {1\over\rho}
\left({\bmsy\nabla}\times{\bmit B}\right)\times{\bmit B} = 0,
\label{forceeq}
\end{equation}
\begin{equation}
{\partial p\over\partial t} + {\bmit u}\cdot{\bmsy\nabla}p
+ \gamma p{\bmsy\nabla}\cdot{\bmit u} = 0,
\label{energy}
\end{equation}
\begin{equation}
{\partial{\bmit B}\over\partial t} - {\bmsy\nabla}\times
\left({\bmit u}\times{\bmit B}\right) = 0,
\label{induct}
\end{equation}
along with the constraint ${\bmsy\nabla}\cdot{\bmit B}=0$ imposed to
account for the absence of magnetic monopoles (e.g., \cite{priest84}).
The entropic gas equation of state is $p\propto \rho^{\gamma}$, where $\rho$
and
$p$ are the density and gas pressure.
Standard symbols are used for other common quantities.  Here, we have
chosen rationalized units for the magnetic field so that the magnetic
pressure $p_b = {1\over 2}B^2$ and the Alfv\'en speed is simply $c_a =
B/\sqrt{\rho}$.

These equations were solved numerically using a multidimensional MHD
code based on the explicit, finite difference ``Total Variation
Diminishing'' or ``TVD'' scheme.  That method is an MHD extension of
the second-order finite-difference, upwinded,
conservative gasdynamics scheme of Harten
(1983), as described by \cite{ryuj95}.  The multidimensional version
of the code, along with a description of various one and two-dimensional
flow tests is contained in \cite{ryujf95}. The code contains an fft-based
routine that maintains the ${\bmsy\nabla}\cdot{\bmit B}=0$ condition
at each time step within machine accuracy.

Numerical solution of equations \ref{masscon} - \ref{induct} on
a discrete grid leads, of course, to diffusion of energy and momentum as well
as
to entropy generation. Of course, such effects are also present
in nature and are important to defining the character of the flows.
The existence of effective numerical resistivity is necessary, for
example, to allow magnetic reconnection to occur.
There is fairly good evidence that conservative
monotonic
schemes, as this one is, do a good job of approximately representing
physical viscous and resistive dissipative processes that are expected
to take place on scales smaller than the grid ({\it e.g.,} \cite{porwood94}).
For the astrophysical environments being simulated the expected
dissipative scales are likely very much smaller than those that can be modeled
directly.
One anticipates then that increased numerical
resolution leads to solutions in which the large-scale flow patterns
``converge'' in a statistical sense over time scales of interest, or
in more detail over limited time intervals.

Our calculations are carried out in the ($x - y$) plane with the
initial background flow aligned with the $x$ direction. For those
simulations presented here we have used a computational space that is
square and extends over $x=[0,L_x]$ and $y=[0,L_y]$, with $L_x = L_y =
L$. In some related tests we applied different proportions for $L_x$
and $L_y$.  The calculations are formally of ``$2~+~{1\over 2}$''
dimensions, since we can include the $u_z$ and $B_z$ components.  In
practice, however, $u_z = 0$ and $B_z = 0$ for the simulations we
present here. We will consider the more general case elsewhere.
Thus, the simulations are practically two-dimensional.  Following MP, the
computational domain is periodic in $x$, while the $y$ boundaries are
reflecting; i.e., both normal velocity and magnetic field change sign
across the top and bottom boundaries.  In addition it is
straightforward to show using equation (2.4) and Stokes theorem that
under these conditions we expect ${{\partial \langle {\bmit B} \rangle } \over
{\partial t}} = 0$, where the brackets represent a spatial
average over the computational domain .  We have confirmed that this
is exactly true in our simulations.  The field is locally changing, of
course, and the mean magnetic energy and pressure are variable, as
well.

We consider an initial background flow of uniform density, $\rho = 1$
and gas pressure, $p = 0.6$, and an adiabatic index, $\gamma = {5\over
3}$, so that the sound speed, $c_s = \sqrt{{{\gamma p}\over {\rho}}}=
1.0$.  The magnetic field, ${\bmit B_o} = B_o \hat {\bmit x}$, is also
uniform, but its value is chosen to be either $B_o = 0.4$ or $B_o =
0.2$ in the two cases we have considered, as described below.  The
velocity in the background state is antisymmetric about $y = L/2$
according to the relation
\begin{equation}
{\bmit u_o} =u_o(y)\hat {\bmit x}  = ~-~{U_o
\over 2} \tanh({y - {L\over 2} \over {a}})\hat {\bmit x},
\label{vprof}
\end{equation}
with $U_o = 1$.
This describes a smoothly
varying flow within a shear layer of full width $2a$.
Flow is to the left in the top half-plane and to the right below that.
To this state we add a perturbation, $\delta (\rho, p, \bmit u, \bmit B)$.
As in MP we define that perturbation
to be a normal mode found from the linearized MHD equations, periodic
in $x$ and evanescent in $y$, with period equal to the
length of the computational box, $L$. Thus, each perturbed quantity can be
expressed in a form,
\begin{equation}
\delta f(x,y,t) = f(y)\exp(ik_x x + i\omega t),
\label{delf}
\end{equation}
where each $f(y)$ is a complex function determined by numerical
integration of the linearized MHD equations as outlined in MP.
Physically, one requires the real part of each $\delta f(x,y,t)$, of course.
Further, we have
\begin{equation}
\label{kxeq}
k_x = {2 \pi \over L}
\end{equation}
and
\begin{equation}
\omega = \omega_r - i \Gamma.
\label{linrat}
\end{equation}
{}From symmetry the real frequency, $\omega_r$, is zero in the
computational frame, so that disturbances remain stationary as they
evolve.  The growth rate, $\Gamma$, can be computed by iteration on
the solution for the $f's$, although we found it adequate to obtain
$\Gamma$ from the published figures in MP. Other parameters that
characterize the system are the Alfv\'enic Mach number, $M_a =
U_o/c_a$, and the sonic Mach number, $M_s = U_o/c_s$.  It is common
and convenient also to use the parameter, $\beta = {{p_g}\over {p_b}}
= {6\over 5} ({M_a\over M_s})^2$, to measure the relative importance
of thermal and magnetic pressures.  For the initial conditions given
above we have $M_s = 1$ and either $M_a = 2.5$ ($\beta = 7.5$) or $M_a
= 5$ ($\beta = 30$).  These properties are summarized in Table 1.  To
minimize the influence of the reflecting boundaries we followed
Miura's (1984) prescription to keep $L \ge 20 a$.  For the simulations
presented here, $L = 25 a$. With this scale boundary zone errors in the
initial normal mode perturbation are insignificant. It is most
convenient in using MP's
formulation to determine the various perturbed variables from a
solution to $\delta p^*$, where $p^* = p + p_b$. Beginning with an
initial value of $\delta p^*$ at one of the $y$ boundaries, one can
integrate to the midplane and then extend the solution to the other
boundary using appropriate even or odd symmetries for the flow
variables around a pair of symmetry nodes along the line of flow
symmetry.  In the two cases presented, we used $\delta p^* (y = L) =
0.001 p^*_o$, ($M_a = 2.5$), and $\delta p^* (y = L) = 0.01 p^*_o$,
($M_a = 5$).  The node points were $y = L/2$, $x = L/4, 3L/4$.  From
the initial conditions (background state + perturbations) we allowed
the unstable flow to evolve until it appeared to approach an apparently
relaxed state.  In practice this meant running the simulations for up
to 20 linear growth times; i.e., $\tau = 20~t/t_g$, where $t_g =
\Gamma^{-1}$.

To understand better how well our solutions are numerically converged
in the sense described near the beginning of this section, we computed each
case on the above
domains using several different resolutions.  Each case was computed
on both $256\times 256$ and $512\times 512$ grids.  In addition the
$M_a = 5$ case was carried out on a $128\times 128$ grid. The highest
resolution simulations required between 20 and 40 CPU hours on a Cray
C90 computer.  Setting $\lambda = {{2\pi}\over k_x} = L$ maximizes our
ability to resolve structures small compared to the initial perturbed
wavelength.  On the other hand, one might be concerned that important
structures should form on scales longer than the initial perturbation
wavelength.  That is not possible when $\lambda = L$, of course.  To
evaluate this concern we repeated the $256\times 256$, $M_a = 5$
simulation on a $x = [0,2L]$, $y = [0,L]$ domain using a $512\times
256$ grid. In other words
we extended the computational box to include two perturbation
wavelengths, but kept everything else unchanged. The results were
indistinguishable from the simulation on the square box, so we conclude that no
important information of
this type has been lost by letting the perturbation wavelength
match the box length.  We should emphasize,
of course, that the choice of periodic boundaries has itself
restricted possible Fourier components, prohibiting the kind of large
scale coalescent structures noted by \cite{wu86}, for example. So, as
pointed out by Wu, a study designed to predict the detailed structures
in an unstable, convected shear layer at an arbitrary stage would need
a different kind of symmetry.

We conducted one additional set of tests to evaluate a related issue; namely
the influence of the location of the reflecting boundaries.  Here we
doubled the distance to the reflecting boundary, $y = [0,2L]$.
In this case we concluded that
the general properties of the flows were unchanged, especially within
the central, strongly sheared region.  Some structural details were
slightly changed, but not sufficiently to alter any of the conclusions
we will present below.

In summary, on a space periodic in one direction we computed the
evolution of two uniform density, but compressible MHD KH unstable
shear layers through their initial exponential growth, saturation and
nonlinear relaxation phases. Both cases had in common an initial
magnetic field aligned with the fluid flow and a sonic Mach number for
the velocity spread, $M_s = 1$.  The two cases were distinguished only
by the Alfv\'enic Mach numbers; those being $M_a = 2.5$ and $M_a = 5$.
Although these Alfv\'enic Mach numbers differ by just a factor 2, the
flows are qualitatively very different.

\section{ Results}

\subsection{Overview: Weak Field and Strong Field Conditions}

Before discussing the two MHD shear layers we have studied it is
useful to outline very briefly the well-known behavior in a
two-dimensional gasdynamic, unstable shear layer (e.g., \cite{mas81}).
For comparison we
carried out a simulation equivalent to the MHD cases described here,
except that the magnetic field was absent.  In that case the
perturbation, once it becomes nonlinear, quickly evolves into a
single, large vortex that extends across the full length, $L$, of the
grid. The vortex is flattened in $y$, so that it spans about $1/3$ of
the grid in that direction. Since numerical viscosity is small, the
fully formed, elongated vortex, sometimes called ``Kelvin's Cat's Eye'',
is stable and would spin almost indefinitely.  It becomes
the only identifiable structure in the flow.  Considering the periodic
nature of the computational space, there would in the gasdynamic
situation be a periodic line of vortices separating the two oppositely
moving and mostly undisturbed fluids.
For nonperiodic flow, vortex coalescence would lead to merging of these
structures (e.g., \cite{corsher84}).
Eventually, over a timescale several orders
of magnitude longer than we have considered, numerical viscosity would
dissipate the vortex flow and the initial kinetic energy would be
converted into thermal energy.  That history, as we shall see, is
rather different from what happens in either of the MHD flows we
studied.

A sense of the evolutionary histories of MHD cases can be obtained
by examining the transverse velocity, $u_y$.  Initially, that is just
the perturbation $\delta u_y$.  Figs. 1a and 1b show the spatial
RMS values, $\sqrt{\langle u_y^2\rangle}$, for both cases as
functions of time, given in units of the predicted growth times; i.e, $\tau
= t\Gamma$.  The different curves represent the various numerical
resolutions used.  Velocities are normalized by the initial value
found for the
perturbation in each run. Several key features are immediately
evident.  First, for both cases the amplitude, $\sqrt{\langle
u_y^2\rangle}$, initially increases exponentially as expected, with a
growth rate close to the theoretical value (shown by the short dashed line in
each panel).  For the $M_a = 2.5$ case there is a startup error that
disturbs the growth momentarily, but it then resumes the predicted
rate very closely. The error, which we attribute to the way our code's
Riemann solver handled the initial conditions at one point along the
midplane, is greater at the lower resolution, and it disturbed an even
lower resolution run ($128\times 128$ grid) sufficiently that we
elected to ignore those results.  The second point to be made from
Fig. 1 is that after only a few linear growth times, $\sqrt{\langle
u_y^2\rangle}$ saturates and then relaxes towards values that are
below the initial perturbed levels. That ``final'', ``relaxed'' state
has some interesting properties, so we shall return shortly to discuss
it in some detail.  The relaxed states that develop are analogous to the
long-lived stable vortex described for gasdynamics.  Since that state
cannot be the ultimate condition for these flows in the presence of
finite dissipation, we will refer to it as a ``quasi-steady'' relaxed
state.  We also note that the time for an initial perturbation
to reach saturation depends on the amplitude of the perturbation.
However, the time elapsed from
saturation, measured in units of $\tau$, ought to be characteristic of
the properties of the unperturbed configuration, including the symmetry.
This last statement comes from the observation that for a given background
flow the characteristic global properties when the perturbation saturates
are not very sensitive to the details of the perturbation itself. This
is demonstrated for the present case by the good agreement between the
saturation-phase flow properties seen in our simulations and those
of \cite{maletal95}.
The latter authors used a simple periodic velocity perturbation rather than a
linear
normal mode.

The third point from this figure is that the flow evolution is
considerably more complex in the $M_a = 5$ case than in the $M_a =
2.5$ case.  The difference in flows is even more obvious in the images
of gas density and magnetic field lines displayed from our highest
resolution runs in Figs. 2 and 3.  To facilitate visualizing the extension of
structures on the periodic space we doubled each image.
For the $M_a = 2.5$ case flows
remain relatively laminar throughout, and qualitatively resemble the
initial, simply periodic form well past the stage when a linear
description becomes invalid.  The magnetic field lines as shown in Fig
2b suffer relatively modest bending or stretching.  There is no
indication of magnetic reconnection, nor that any suitable field
topology for that process has formed.  Once the oscillations reach their
maximum
amplitude, the flow seems simply to flatten out into a broadened shear
layer.  In marked contrast, the $M_a = 5$ case develops flows that
become quite complex, with distinct, intermittent vortices.
The large amplitude
oscillations in $\sqrt{\langle u_y^2\rangle}$ after $\tau \approx 2$
are associated with the generation and decay of these vortex
structures. The large ``primary vortex'' centered on $x = 3L/4$ at the
$\tau = 2.4$ in Figs. 3 reaches its maximum development after $\tau =
2$ and corresponds to the first maximum in $\sqrt{\langle
u_y^2\rangle}$.  Around $\tau = 6$ the magnetic field lines on the
outside of that vortex, having been wrapped back on themselves, tear
in a major reconnection event that matches with the second maximum in
$\sqrt{\langle u_y^2\rangle}$.  Fig. 3b shows the field configuration
at $\tau = 5.6$, just before that event.  The third maximum in
$\sqrt{\langle u_y^2\rangle}$ corresponds to a similar reconnection
event associated with the well-formed ``secondary vortex'' centered at
$x = L/4$ and $\tau = 8$ in Fig. 3.  Smaller peaks in $\sqrt{\langle
u_y^2\rangle}$ can also be seen at $\tau \approx 12, 15~{\rm and}~
18$. Each is associated with reconnection in a vortex structure within
the flow.

There are clear differences between the descriptions just given for
the MHD flows and that presented earlier for the analogous gasdynamic
situation.  Even when the magnetic field is initially weak its
presence becomes crucial in determining the nonlinear evolution
of the flow. The initially weak field is able to mediate dynamical and
dissipative processes through local growth and decay.
Not surprisingly, the role played early on by the magnetic
field depends considerably on its initial strength. That is apparent in the
initial, exponential growth rates.  The growth rate for the $M_a = 5$
case ($\Gamma = 0.108 \times {{U_o}\over {2a}}$) is about 90\% of the
purely hydrodynamic rate ($\Gamma = 0.122\times {{U_o}\over {2a}}$
\cite{miupr82}), whereas the rate for the $M_a = 2.5$ case ($\Gamma =
0.053\times {{U_o}\over {2a}}$) is less than half the (same)
hydrodynamic rate.  From these various contrasts it is convenient and,
we believe appropriate, to distinguish in our remaining discussion the
two cases as ``weak field'' ($M_a = 5$) and ``strong field'' ($M_a =
2.5$), thus being generally representative of qualitatively different
flows.  \cite{maletal95} also saw a qualitative transition in
properties between these two field strengths.
We note for comparison that the incompressible hydrodynamical
linear perturbation growth rate of a discontinuous shear layer
is $\Gamma = {\pi}\times {{U_o}\over
{\lambda}}$.  Here, $\lambda = 25 a$ ($\Gamma \rightarrow
{\pi\over{12.5}}\times {{U_o}\over {2a}}$), so a combination of
compressibility and the finite spread in the shear layer has reduced
the initial growth rate by more than a factor of two even before Maxwell
stresses are introduced.  The sonic Mach number is unity, after all,
and in each case thermal energy is globally the dominant energy form
throughout the simulations by about an order of magnitude.

As noted before, numerical resolution is an important issue in
evaluating computations of this kind.  We will comment on specific
resolution-related questions as they come up, but we can make some
preliminary observations here.  It is clear from an examination of
Fig. 1 and by comparing images of various physical quantities, that
many qualitative behaviors are similarly captured in all of the
calculations.  In Fig. 1b, for example, the same major oscillations in
$\sqrt{\langle u_y^2\rangle}$ occur for all three weak field
simulations. By viewing animations of the three flow simulations one
can see that the oscillations represent the same flow events as
well. Likewise, there are no essential differences in the global
characters of the two strong field simulations.  At the same time it
is obvious that the lower resolution runs are not converged
quantitatively in important physical variables such as peak magnetic
energy, or kinetic energy during the saturation phase.
\cite{ryujf95}
demonstrated with this code that under conditions similar to the
strong field case here, isolated structures need to be resolved
within about 30 zones before one can realistically neglect numerical
diffusion; i.e., before the effective ``Reynolds number'' of the structures
$> 10^3$.  For conditions
similar to the weak field case that requirement is somewhat less
restrictive. There we can estimate that about 20 zones are
needed. Thus, while our lowest resolution runs would contain as few as
6 ``fully resolved'' structures in each direction, that
number climbs to around 25 or more in the highest resolution
simulations.

We will now proceed to discuss the characteristics of three phases in
the nonlinear evolution of the simulated flows. From the preceding
outline we identify these as: a) saturation, b) relaxation and c) the
quasi-steady, relaxed state.  The initial, exponential growth phase is
simply an extension to
larger amplitude of the initial perturbation as discussed in MP, so
we will not discuss it here.  The quasi-steady
relaxed state, is relatively simpler than any of the other nonlinear
stages and comes closest to having conditions that are straightforward
to predict.  Consequently, we begin our analysis there.

\subsection{The Quasi-steady, Relaxed State}

As mentioned in the previous summary, the simulated flows in both
strong and weak field cases relax fairly quickly to an apparently
steady condition.  The fact that we have used periodic boundaries in
our simulations and have conserved total energy in the domain of
computation means that eventually the flow kinetic energy must be dissipated.
But, that ultimately anticipated condition is not what is observed
here, as emphasized by the comparison made with the gasdynamic case.
The final density and magnetic field structures in our MHD simulations
as shown at $\tau = 17.9$ for the strong field flow in Fig. 2 and
at $\tau = 20$ for the weak field flow in Fig. 3 give a sense of
the character of the relaxed states reached.  The flows are nearly
laminar, reflecting the fact that the mean transverse velocity has
decayed to very small levels by this time (see Fig. 1). There are
remaining visible striations in both the density and magnetic fields,
however (see also Fig. 4).  Within the magnetic field variations the
magnetic and velocity fields are almost perfectly aligned, so that the
normalized cross helicity,
\begin{equation}
\langle H \rangle =
{{\int \hat {\bmit u} \cdot\hat {\bmit B}dxdy}\over{\int dxdy}},
\label{crosshel}
\end{equation}
very closely approaches the possible extrema, $\mp 1$, in the top and bottom
half planes
respectively. Thus, the flows are highly organized by this
time. Initially, the cross helicity is also large, but for the weak
field flow it becomes much reduced during the nonlinear phases of flow
evolution. In that case the maximized cross helicity at the end may be
related to so-called ``dynamical alignment'', first noted with respect
to similar characteristics in the solar wind by \cite{dobro80}, and
subsequently examined theoretically and numerically in more general
contexts of MHD turbulence (e.g., \cite{pouq86,tingetal}).  Outside of
the small influences of the striations, the transverse velocity
gradients, $\partial u_x/\partial y$, are almost constant within the
broadened shear layer. We have examined animations of a number of
variables to ascertain any remaining trends in the flow
evolution. Although it is apparent that some of the feature details
may be expected to relax further, it seems clear to us that the
general properties of the flow patterns visible at these times will
continue to exist as seen well beyond the end points of our
simulations and that no new features are developing.

The properties of the magnetic field in the relaxed states are perhaps
the easiest to understand.  They come largely from the assumed
symmetry in the problem and the laminar nature of the flow in the
relaxed states.  $\langle B_x \rangle$ and $\langle B_y \rangle$ are
both exactly conserved during the calculation, so in the relaxed
states we may expect fairly uniform fields.  The magnetic energy in
the flow may be reasonably expected to return to something resembling
its original value.  The manner of those trends can be made evident by
using equation (\ref{induct}) to derive a relation for the evolution in
the mean magnetic energy density, $\langle E_b \rangle = {1\over 2}
\langle B^2 \rangle$.  As mentioned in the introduction we depend on
numerical dissipation to mimic the
effects of viscosity and resistivity, although we cannot describe
an analytic model for the two effects or accurately separate them.
However, to understand heuristically their effects
qualitatively we can add a term $\eta \nabla^2 \bmit B$ to equation
(\ref{induct}), where $\eta$ represents the effects of numerical
resistivity. For a two-dimensional flow with our boundary conditions
this leads to the equation
\begin{equation}
{{d \langle E_b\rangle}\over{dt}}={\partial \langle E_b\rangle \over
\partial t} + \langle ({\bmit u \cdot \nabla})E_b\rangle = -\eta
\langle{\bmit j}^2\rangle - \langle 2E_b{\bmit \nabla \cdot u}\rangle
+
\langle {\bmit B} \cdot [{\bmit B \cdot \nabla}{\bmit u}]\rangle,
\label{ebent}
\end{equation}
where ${\bmit j} = {\bmit \nabla \times B}$ is the current density,
and we assumed $\eta$ to be constant.  The average is over the entire
domain.  The three right hand terms represent respectively, Joule
dissipation or magnetic diffusion, compression of the field and field
line stretching.  Unless the relaxed field configuration is uniform
there will ordinarily be a finite $\langle {\bmit j}^2 \rangle$, and
hence dissipation. That term always will decrease the magnetic energy
and can only be countered by enhancement through compression and
stretching.  But, in the laminar relaxed state those effects are
fairly small, because of the absence of strong gradients in the
velocity structure.  Thus, we must expect that the final field will be
relatively uniform and that the associated magnetic energy will
resemble the initial value in order to satisfy the constant mean field
values.

Fig. 5 shows the fractional change between the initial and final
magnetic energies, $\Delta E_b/E_{bo}$, for both the strong and weak
field cases as a function of the number of grid points, $N_x$, used in
the various simulations.  We see that for the strong field case the
final and initial magnetic energies are, indeed, almost precisely the
same for both numerical resolutions used.  They agree to much better
than 1\%.  For the weak field case, the two lower resolution
simulations also relax to a state with magnetic energy equaling the
initial value to better than 1\%. Remarkably, however, the highest
resolution simulation in this case ends with a magnetic energy greater
than the initial value by about 7.7\%.  Fig. 6, which shows the time
evolution of $\langle B^2_x\rangle$ and $\langle B^2_y\rangle$ for
that simulation, also illustrates this fact.  Although $\langle
B^2_y\rangle$ returns very close to its initial value,
$\langle B^2_x\rangle$ does not evidently do so. We cannot rule out
the possibility that $\langle B^2_x\rangle$ will slowly decay to its
initial value, but we prefer another interpretation of this result;
namely, that it represents the initial development of
``two-dimensional flux
tubes''.  Fig. 4 shows that the final state in this flow contains a
pair of broad striations in enhanced magnetic pressure positioned on
opposite sides of the midplane, $y = L/2$. It is also apparent from
Fig. 4 that the magnetic and gas pressure distributions are strongly
anti-correlated at this time.
Moreover, the maxima in magnetic pressure are also
local minima in gas density. These features can, therefore, be
described as slow mode rarefactions, similar to the ``plasma
depletion'' zone described for an earlier stage of the flow by
Miura. In addition, we already mentioned that the velocity and
magnetic field vectors are almost exactly parallel in the lower half
plane and anti-parallel in the upper half plane. Thus, the magnetic
field lines and velocity streamlines are aligned and the flow is
directed along the tubes.  By watching animations of the evolution of
the magnetic field structure one can see that the two flux tubes
(stronger field regions) have survived as coherent features since the
reconnection event associated with the breakup of the large vortex.
The tubes can be identified relatively easily as clusters of
field lines stretching obliquely above and below the vortex
in the $\tau = 8$ panel of Fig. 3b. From that time
they gradually settle down towards the midplane as activity in that
region decays.  We see no evidence of these tubes as coherent
structures in our lower resolution simulations of this flow.
Numerical diffusion presumably prevented them from forming, a fact
that emphasizes the need for adequate numerical resolution in
simulations of this type.  There is also no indication of such
structures in the strong field case, because there were no
opportunities for dynamical self-reorganization through magnetic
reconnection during evolution of that flow.

These final structures are also apparent in the Fourier power spectrum of
the magnetic pressure, $p_b$.
We constructed two-dimensional power spectra, $P^b_{x,y}$,
at a number of times for both the magnetic field cases.
The power spectra are not isotropic, as obvious from the flow structures
seen in Figs. 2 and 3.
Initially, $P^b_x$, the power spectrum along different $k_x$, has
a single mode at $k_x = 1$ of course.
$P^b_y$, the power spectrum along different $k_y$,
has a fairly broad distribution
at the beginning, reflecting the finite thickness of the shear layer.
At the end of the simulations, $P^b_x$ has returned to have almost
a monochromatic mode for the both cases.
For the strong field case $P^b_y$ is also similar to its initial form,
but with two notable changes.
The dominant mode is now at $k_y = 3$ rather than at $k_y = 1$.
In addition, it is spread over more modes to reflect an increased
thickness of the shear layer, and there is some structure in the
power spectrum coming from the features that can be seen in Fig. 2.
In the weak field case $P^b_y$ has a narrower peak than that at the beginning
with the dominant power at $k_y = 4$, but also with distinct peaks at
$k_y = 8$ and 12.
Some power extends to the highest modes in this case, as
expected from thin stripes seen in Fig. 3.

As already mentioned, except for small deviations aligned with
remaining magnetic features, the final velocity structure can be
described as a simple shear layer (see Fig. 3c).  During the evolution
of the flow the initial shear layer spreads, of course, because
momentum is transported between the bottom and top halves of the
computational domain.  Horizontal momentum, $\rho u_x$, is transported
in the $y$ direction by way of Reynolds stresses, $\rho u_xu_y$,
Maxwell stresses, $B_xB_y$, and also by numerical diffusion.  In the
same spirit as we had for equation (\ref{ebent}), we can represent that
last behavior qualitatively in terms of an effective viscous stress,
$\sigma_{xy}$, (e.g., \cite{lanlif87}). Then, for example, using the
symmetry and boundary properties of this problem the change in the
mean $\rho u_x$ within the {\it upper half} of the domain is
\begin{equation}
{{\partial \langle \rho u_x\rangle}\over {\partial t}} = {2\over L^2}
\left\lbrack\int^{L}_{o} (\rho u_x u_y - B_x B_y + \sigma_{xy})dx
\Bigg|_{y = {L\over 2}}\right\rbrack
\label{momtran}
\end{equation}
(see also \cite{miura84}).  The shear layer should spread so long as
$\partial \langle \rho u_x \rangle /\partial t > 0$, since initially
$\langle \rho u_x\rangle < 0$ in this region.  Once the flow becomes
nearly laminar the Reynolds and Maxwell stress contributions to
momentum flux across the midplane should become small, leaving only the
viscous stresses as a source of transport. Although viscous stresses
can never be entirely eliminated, they can be reduced to nominal
levels as the numerical resolution is increased.  For our highest
resolution runs of both cases the momentum flux as measured by the
right hand side of equation (\ref{momtran}) is very small for $\tau > 10$.
It is almost exactly zero in the strong field case, but oscillates
around zero with decreasing amplitude in the weak field case.

As the disturbed flows become nearly laminar, shear layer spreading
slows for reasons just given. We will address in \S 3.3 how that
return to near laminar flow happens. But, one might reasonably ask why
remaining fluctuations in $u_y$ do not lead to renewed growth and
further spreading as that final state is approached. Qualitatively, at
least, the answer seems straightforward.  In particular, if we note
that the $u_x$ structure of the final state is similar to that in the
initial flow except for the width in $y$, then we know from the linear
dispersion relation (e.g., MP) that there is a minimum unstable
wavelength related to the thickness of the shear layer; i.e., for
unstable perturbations, $k < {\phi\over {2a}}$, where $\phi \sim
1~-~2$, depending on $M_a~{\rm and}~M_s$.  The fastest growing modes
are concentrated at wave numbers about half this limit.  On the other
hand, our periodic boundary conditions limit Fourier components in the
flow to $k\ge {2\pi\over L}$.  Thus, we expect that once the shear
layer has spread so that the nominal width, $a\sim |U_o/(\partial
u_x/\partial y)| > {{\phi L}\over{4\pi}}$, a quasi-laminar flow will
be mostly stable to any further perturbation, independent of the
magnetic field strength.  For both cases we studied the final shear
layer width as estimated from the slope of the central region is at
least $\sim 0.5 L$. So, from the conditions just derived, we would not
expect renewed growth.  For the weak field case this estimated final
value of $a$ from the observed flow is $\sim L$, which could be
limited or modified by boundary effects. As further confirmation of
our conclusion, then, we carried out a $256\times 256$ simulation
using $L_x = L_y = 2L = 2\lambda$.  Except for details in the flows
near the top and bottom boundaries the properties of that flow were
the same as for the $128\times 128$ simulation using $L_x = L_y = L =
\lambda$.

We previously noted that in the relaxed state the magnetic field lines
and flow streamlines are highly correlated.  That is evident in
Fig. 3b and 3c.  In fact it is apparent that within the ``flux tubes''
velocity and magnetic field fluctuations satisfy the relation $\delta
{\bmit u} \sim \mp \delta {\bmit B}$, where the minus sign corresponds
to the fluctuations in the upper half plane and the plus sign
corresponds to the lower half plane.  That relation is consistent with
the remaining fluctuations being carried in Alfv\'en waves that
propagate locally downstream within the flux tubes and are linearly
polarized so that $\delta u_z~=~\delta B_z~=~0$. This feature and the
convergence and divergence of flow along the varying width of
the tubes explains the good correlation of
magnetic pressure and $u^2$ observed when the velocity is corrected
for the mean shear structure, $\langle u_x(y)\rangle$.

In contrast to the magnetic energy, we should anticipate that the
final total kinetic energy inside the computational domain, $E_k$,
will be reduced in response to the relaxation of the flows.  Since in
the relaxed flow the velocity transition is spread out more between
the top and bottom states, and the computation was carried out in the
center-of-momentum frame, we can expect that the quantity $\langle
u^2_x\rangle$ is reduced over the domain by the spreading of the
transition.  In both the strong and weak field cases the kinetic
energy in the relaxed state exceeds the magnetic energy. For the
strong field case $E_k/E_b \approx 1.09$ at the end of the
simulation. It is conceivable that an eventual fairly long term
equipartition between magnetic and kinetic energy develops.  However, kinetic
energy in the weak field flow always significantly exceeds magnetic
energy in our simulation. In the relaxed state $E_k/E_b \approx 2$.
On a much longer time frame, it is not inconceivable that an equipartition
between kinetic and magnetic energies develops in response to the
dissipative decay of the flow mentioned in the next paragraph.  On the
other hand, it is also possible that the observed non-equipartition condition
{\it does} represent the relaxed condition of the flow.
\cite{tingetal} have pointed out that turbulent MHD flows dominated by
dynamical alignment can trend to long-term states in which the kinetic
energy remains dominant over the magnetic energy.

Since the magnetic energy is little changed at the end from the
beginning and kinetic energy is reduced, it follows that the final
states must possess increased thermal energy over the initial flows.
Figs. 7 and 8, which display the evolution of the total kinetic,
magnetic and thermal energy forms over time, verify this.  For the
strong field case the kinetic energy decreases by almost 24\%, while
that number is close to 60\% for the weak field case.  This leads to
increases of thermal energy, $E_t = p_g/(\gamma - 1)$, of about 3.3\%
(7.8\%) in the strong (weak) field case by the end of our
simulation. Taking advantage of the symmetries and boundary properties
in the simulations, it is straightforward to show that the mean
thermal energy obeys the relation
\begin{equation}
{{\partial \langle E_{t}\rangle}\over{\partial t}} = - \langle \ p_g
{\bmit \nabla \cdot u}\rangle +\eta \langle {\bmit j}^2 \rangle +
\langle \sigma_{ik} {{\partial u_i}\over{\partial x_k}}\rangle,
\label{etherm}
\end{equation}
where we have once again for heuristic purposes included effective
resistive and viscous dissipation terms in addition to the
consequences of adiabatic compression. Below we use equation
(\ref{etherm}) to estimate the degree to which our simulations
are influenced by finite numerical resolution.

In these calculations the total mass is conserved inside the grid.
Furthermore, the use of periodic boundaries means that any adiabatic
work applied to mass leaving the domain is exactly replaced by mass
with identical properties
entering from the opposite boundary; i.e., thermodynamically, the
system is closed. Thus, since the average compression must be zero
over the fixed volume, the net adiabatic work done on the gas is expected
to be small, and the first term in equation (\ref{etherm}) is small.
Consequently, ${{\partial\langle E_t\rangle}\over{\partial t}}$ must
be due almost entirely to the two dissipative effects.  We have verified
numerically that $\langle \ p_g {\bmit \nabla \cdot u}\rangle$ is indeed
small.  There are no shocks formed in either of these cases, so only
dissipation in smooth flows matters.  Even though the dissipation in
our computation is numerical, that does not necessarily mean the
simulated flows are physically incorrect, of course, at least in a
statistical sense.  Pseudo-ideal fluid
simulations of shocks and turbulence, for example, depend on the
existence of a small amount of numerical dissipation
to obtain physically meaningful
behaviors. But, it is necessary that the numerical dissipation be
small enough that the computed flows approach the intended physical
flows (meaning those with very small viscosity and resistivity).  To
evaluate this situation we should ask how well the total dissipation in the
simulated flows has converged. Examination of the time evolution of
the thermal energies as shown in Fig. 7 and 8 provides that
information.  The final increases in thermal energies agree among the
different simulations of the same flows to within 10\% of the same
value in the strong field case, and about 6\% for the weak field.
Therefore, we consider the final states moderately well converged in
that respect.  We will explore in the next subsection some details
about how that dissipation takes place prior to the development of the
relaxed states.

We emphasize again that, although the final conditions of our
simulations are not changing very fast, they cannot represent the
expected conditions if one extended the computations for an
arbitrarily long time.  In fact, dissipation cannot entirely vanish so
long as there is any shear left within the system.  In the limit that
the flow is laminar the viscous term in equation (\ref{etherm}) can be
represented qualitatively in terms of an effective shear viscosity,
$\mu$, and the mean-squared vorticity or enstrophy in the flow,
$\langle W \rangle = \langle ({\bmit \nabla \times u})^2\rangle$, as
$\langle \sigma_{ik} {{\partial u_i}\over{\partial x_k}}\rangle \sim
\mu \langle W \rangle$. From this and equation (\ref{etherm}) the
time for the flow kinetic energy to decay to thermal energy is $> R
a/U_o$, where $R$ is the effective viscous Reynolds number in the
flow and $a$ is the nominal width of the shear layer. Using the properties
of our code as determined by \cite{ryujf95}
we can estimate that $R \gsim 10^5$ in our highest resolution runs.
So, this decay time is several orders of magnitude longer than our
simulation.

We noted earlier that the final magnetic field configuration in the
weak field case showed signs of an organized structure; namely flux
tubes containing magnetic field lines aligned with the flow
streamlines.  The final state of the weak field flow contains a second
structure that warrants comment.  Examination of Figs. 3a and 4 shows,
in addition, the presence of a low density ``channel'' containing hot,
high entropy gas concentrated along the midplane.  That feature does
not correspond to the flux tubes, but rather lies between the flux
tubes. The origin of this second feature becomes obvious if one views
animations of the evolution of several quantities, especially the
density, magnetic field, electric current and entropy, $S =
\ln(P/\rho^{5\over 3})$.  Most of the entropy generated in dissipative
events that we will describe in the next subsection has been
concentrated into this thin layer, which appears to be contained by
the two magnetic flux tubes, since they lie just outside it.  Gas
within the central layer is about 25\% hotter than the average over the
domain.  The gas pressure is moderately high within the layer, but
there is a near equilibrium balance in the total pressure, $p~+~p_b$.
Recalling that the flux tubes were traced to an early magnetic
reconnection event separating the magnetic field in the outer and
inner parts of the flow, we can recognize that a consequence of the
evolution in the weak field KH unstable flow has been a reorganization
of the flow into two distinct regions; one in which the energy
dissipation necessary for relaxation takes place and one that surrounds
that region.  There is no evidence of such a development in the strong
field flow, again presumably because the stronger initial field
prevents the opportunity for the reorganization.

\subsection{Saturation and Relaxation}

As noted before, the history of a simulated two-dimensional
ideal gasdynamic shear
layer in a periodic flow is very simple; i.e., a single,
long lived vortex forms.  That
history seems quite distinct from what happens in the MHD flows, where
the formation of vortices is either intermittent or entirely absent.  In
the weak field case we computed the primary vortex begins development,
but is disrupted before it is fully formed. Its destruction leads to
the generation of a sequence of smaller vortices (each located at one
of the symmetry or node points in the computational plane), but they, too, are
quickly destroyed.  In the strong field case the initial oscillation
is stabilized before a vortex forms, and the flow then directly begins
to approach a laminar shear layer. It might seem odd that our two MHD
flows are so different, because their properties as given in Table 1
are the same except for a factor two difference in the strength of the
magnetic field. In both MHD flows the Alfv\'en Mach number is greater
than unity and the plasma $\beta$ parameter is substantially greater
than one.  However, it should be recalled that a parallel flow
discontinuity is KH stable whenever the Alfv\'en Mach number, $M_a <
2$, only slightly smaller than the value for our strong field case.
This criterion represents the condition that the restoring force from
magnetic tension due to field line stretching along a perturbed
boundary equals the ``lift'' force on the boundary.  Thus, for our
strong field flow the initial Maxwell restoring force is only slightly
less than needed to stabilize the oscillation.  By comparing the
magnitude of the Maxwell stress, $|({\bmit j \times B})_y|$, and the
gas pressure gradient, $|\partial p/\partial y|$, in the strong field
flow we confirmed directly that the initial perturbation grows only
until the Maxwell stresses {\it along the displaced boundary} exceed
the gas pressure gradient that drives the instability. From that
moment, which is simultaneous with the peak in
$\langle\sqrt{u^2_y}\rangle$ in Fig. 1a, the two stresses seem quickly
to achieve a close balance almost everywhere, while both transverse
motions and excess magnetic energy steadily decay (see Figs. 1, 2, 3
and 7).  Magnetic enhancements during the growth of the instability
are concentrated into a close pair of
magnetic flux tubes formed along the displaced flow boundary that have
reduced plasma density and pressure. Although the total Maxwell stress
does locally exceed the gas pressure gradient at saturation,
the $\beta$ parameter
is never less than about 2.5 (see Fig. 7), so an ``equipartition'' in
the common sense never develops. The total magnetic energy is enhanced
by about 8\% at its peak in the higher resolution simulation for this
case. That compares to about a 9\% enhancement reported for this case
by Miura, so the agreement is reasonably good.

The early development of the weak field flow is very different, of
course, and resembles in most ways the behavior of the pure
gasdynamical KH instability.  Just as in the gasdynamic case the
applied oscillation grows in amplitude and then begins to role up
into a large vortex, as shown in Fig. 3.  However, it is well known
that the presence of even an initially weak background magnetic field
can have a profound influence on turbulent or sheared flows (e.g.,
\cite{chan60,biswel89,catvain91,balhal91,noret92}), especially if it has a
non-vanishing mean vector value.
That property is soon apparent in this simulation. Instead of
rolling up and continuing to spin, the vortex formed in our weak field
flow becomes distorted in response to increases in magnetic field
strength. Both magnetic pressure and tension contribute. That effect
is apparent at $\tau = 5.6$ in Fig. 3.  The maximum total magnetic
energy during the simulation occurs near this time. That energy, which
exceeds the original value by $\sim 140$\% in the highest resolution
simulation of this case (see Fig. 8), is concentrated into a thin flux
tube feature.
Our peak magnetic energy enhancement in this simulation agrees well
with that found by \cite{maletal95} in their analogous calculation.
The flux tube feature corresponds to the low density plasma
channel that approximately traces the displaced initial shear layer
being wrapped into the vortex at this time. Miura pointed to
the almost straight segment of that feature, centered on $x = L/4$ and
called by him the ``plasma depletion zone'', as the place where most
of the enhanced magnetic energy is concentrated.  However, we find a
much larger total magnetic energy enhancement (his was only 26\%) and
that by far the greater portion of the energy is held within the main
vortex structure itself.  In his computation, which had much lower
resolution and, hence, greater diffusion and dissipation, the vortex was
more strongly damped this time. We find,
as well, a somewhat greater concentration of magnetic field within the
plasma depletion zone, leading to a minimum $\beta = 0.55$, compared
to the $\beta = 1.7$ that he quotes. We note that the plasma depletion
zone seen at this time is {\it not} the same structure that leads to the
two-dimensional
flux tubes described for the final state. Those form separately and
can be seen simultaneously with this feature at $\tau = 8$ in Fig. 3b.

About the time, $\tau \approx 4.5$, that the magnetic energy is
reaching its maximum, the total kinetic energy is at a local minimum
(Fig. 8), a development also apparent in the first local minimum in
$\langle\sqrt{u^2_y}\rangle$ seen in Fig. 1b. Thus, it is clear that
magnetic stresses have impacted on the gasdynamics in a substantial
way. Immediately after these developments, Fig. 8 shows that the
magnetic energy drops, while the kinetic energy rises.  These further
developments come from the fact that as the magnetic field is being
drawn into the vortex, field lines from just outside the initial shear
layer become folded back on themselves around the ``tips'' of the low
density channel (see Fig. 3b).  The magnetic energy maximum
corresponds to the moment just before the field
becomes unstable to a tearing mode
instability near the ``tips''.  That leads to magnetic field
reconnection and field annihilation.  Blobs of heated, low density
plasma are ejected from the ``tips'' into the outer regions of the
flow and the magnetic field reorganizes itself.  The ejected blobs
contain magnetic flux islands; i.e., closed flux loops.  Within the
vortex the field lines snap back towards the vortex center
(accelerating plasma with them as they do so), while the exterior
field structure becomes isolated from further involvement with vortex
behaviors within the central shear layer. The vortex collapses subsequently
into the central shear layer.

As the original, primary vortex is destroyed in the manner just
described, a strong, secondary vortex spins up, centered at $x = L/4$.
That vortex also wraps and amplifies magnetic field within it,
until at $\tau \approx 7.5$, field lines have been pulled completely
around the outside of the vortex causing another reconnection event
(see also Fig. 3). Once again, this event corresponds to a maximum in
the total magnetic field energy. There is also a peak in the kinetic
energy near this time, but this actually corresponds to the
``ejection'' of hot plasma from the primary vortex, at $\tau \approx
6$, when the magnetic energy is minimum. Magnetic energy and vorticity
are concentrated around the perimeter of the vortex. That is also
where reconnection begins. The reconnection event, as well as the one
mentioned earlier, shows up clearly in images of such quantities as
electric current, entropy and $\beta$. Where magnetic ``$\times$'',
or neutral points form, intense local current sheets develop and excess
entropy is generated, along with expansions in response to the
heating.  Local maxima in $\beta$ also develop around the $\times$
points, of course.
As with the earlier vortex-directed magnetic reconnection, the
initial reconnection event here separates magnetic field within and outside
the vortex.  In this case the magnetic field is
somewhat weaker than before and the vortex is stronger.  The vortex
continues to spin for a while and the magnetic islands formed end up
inside the vortex.  It is a magnetically isolated structure, with
detached fields lines surrounding it.  The vortex and the magnetic
energy within it decay by about $\tau \approx 10.5$.  Remnant magnetic
islands are still visible at $\tau = 10.4$ in Fig. 3b.  Weaker
vortices briefly form and decay at later times, leaving noticeable
signature peaks in $E_b$ around $\tau = 14 {\rm~ and~} \tau = 17$.
Those features are anticorrelated with kinetic energy maxima at $\tau
= 15$ and $18$.

We commented earlier on the properties of the power spectrum of the
magnetic pressure structure at the end of our simulations.  To complete that
discussion we briefly add a few comments about the power spectrum
evolution during the saturation and relaxation phases for the weak
field case.  There is little change in the $P^b_x$ spectrum, except that
as the secondary vortex strengthens it leads to a shift from dominance
by the $k_x = 1$ mode to similar strength in both the $k_x = 1$ and
$k_x = 2$ modes. For $P^b_y$ the evolution is more remarkable in that
power spikes at a number of scales are evident. The number and strength
of those spikes changes as the number and scales of the dominant
structures change. For example, as the primary vortex forms there is a
single broad peak, around $k_y = 3$, which breaks up into at least
four comparable narrow spikes at such wave modes as $k_y = 2,~7,~9$
and 13, once that physical feature is destroyed. Those are replaced
during the life of the secondary vortex by a pair of comparable spikes
at $k_y = 1$ and 7. The structure at $\tau = 14.4$ (see Fig. 3)
leads to four comparable sharp power spectrum spikes at $k_y = 1,~4,~6$ and 9.

Since the secondary vortex is strong and spins several times before
decay it presents a good opportunity to study the interactions of the
flow and the magnetic field within it.  The equation for the evolution
of vorticity in two-dimensional ideal MHD can be written as
\begin{equation}
{d\rho {\bmit \omega} \over dt} = ({\bmit \nabla \times (B \cdot \nabla ) B})
+ {\bmit \nabla}\rho {\bmit \times} ({d {\bmit u} \over dt}),
\label{vordif}
\end{equation}
with ${\bmit\omega} = {\bmit\nabla\times u}$.
The ${d{\bmit u}\over{dt}}$ term in equation (\ref{vordif}) represents
the total acceleration of a fluid element due to all forces (see equation
(\ref{forceeq})).  For an
entropic equation of state ${\bmit \nabla} \rho {\bmit\times \nabla}
p = 0$, so the only nonvanishing contributions (for ideal
flows) to the last term would come from the Lorentz force; i.e.,
${1\over\rho}{\bmit \nabla}\rho
{\bmit\times}\left({\bmsy\nabla}\times{\bmit B}\right)\times{\bmit B}$.
On the other hand, these flows are only slightly compressible ($M_s
\sim 1$), so that $|{1\over\rho}{\bmit\nabla}\rho|<<1$, and this term
ought to be small.  Then with the aid of Stokes theorem we can relate
the change in the total vorticity within an area to the line integral
of the magnetic tension around the boundary,
\begin{equation}
{\int {d (\rho {\bmit \omega}) \over dt}\cdot{\bmit dA}} \approx
\oint ({\bmit B \cdot \nabla) B\cdot dl} = \oint {\bmit T \cdot dl}.
\label{inteq}
\end{equation}
Fig. 9 displays the two sides of equation (\ref{inteq})
computed from the flow properties
during the life
of the secondary vortex, where the area integral was carried out over
the vortex and the line integral surrounded it.  The vorticity is
positive inside the area, so the signs indicate a reduction in the
strength of the vortex.  Clearly the two quantities are in good
agreement, so that equation (\ref{inteq}) properly represents the
evolution of the vortex.  Magnetic tension around the vortex peaks in
magnitude at $\tau = 7.5$, in agreement with the magnetic energy peak
noted earlier. Between $\tau = 7$ and $\tau = 9.5$ the value of
$\rho w_z$ inside the vortex decreases from its maximum value
by about 75\%. The integrated
effect of magnetic tension according to equation (\ref{inteq}) accounts
for this change to within about 4\%.

Finally, we make a few comments about the evolution of the cross
helicity in the weak field flow.  The cross helicity, $\langle H
\rangle$, (defined in equation
(\ref{crosshel})) provides a way to express the self-organization of the flow.
We already commented that at the end of our weak field simulation the
cross helicity approaches very close to its possible extrema, $\pm 1$, in the
bottom and top halves of the computational domain.  The two signs just
reflect differences in the initial velocity signs in the two regions
of the grid. Since we began with a condition $\langle H \rangle = \pm
1$ in the appropriate halves, it may not be very surprising that the
flows eventually return to that condition. In fact, if the flows were
exactly laminar, this would necessarily be so.  On the other hand, the
flows still contain fluctuations at the end, and within those
fluctuations the velocity and magnetic field structures are very well
aligned.  At intermediate times the cross helicity is only about half
these extreme values, as can be seen in Fig. 10. So, we believe this
is an indication of real self-organization.  Evidence for
dynamical alignment can be found by comparing Fig. 10 with Fig. 6,
which follows evolution of the mean squared magnetic field
components. The initial large drop in $|\langle H \rangle|$ relates to
the growth of the initial, primary vortex. We have already pointed out
that the valleys in $\langle B^2_x\rangle$ and $\langle B^2_y\rangle$
correspond to reconnection events associated with various flow
vortices.  Each large increase in $|\langle H \rangle|$ visible in
Fig. 10 at, for example, $\tau = 7$, is generally simultaneous with
one of those reconnection events, as well.  Thus, the reconnection process
serves to realign the flow and field vectors as the flow relaxes.

\section{Summary and Conclusion}

We have carried out numerical simulations of the evolution of unstable
sheared MHD flows in two-dimensions. Although we chose to conduct
these experiments in a rather idealized setting, using a box that was
periodic, we believe they have enabled us to discover several
important insights about the nature of such flows that can be helpful
in understanding more general situations. Our unperturbed initial
conditions involved smooth shear layers of constant gas density
separating flows with relative motions at sonic Mach number unity.
There was a uniform magnetic field aligned with the flow.

We considered two cases. In the first the magnetic field strength was
only slightly less than required to stabilize the perturbation.  In
the second case the magnetic field had a value 2.5 times smaller than
that critical value. We followed each flow until it reached an
apparently steady condition.  For both flows the properties of that
steady condition were fundamentally different from what happens in an
analogous, two-dimensional gasdynamic unstable sheared flow. In the
gasdynamic situation the flow evolves to include a single well-formed
vortex that separates the two counter-flowing fluids. That vortex
would spin for a very long time, until viscous interactions spread it
and dissipated it.  By contrast, in the MHD flows, the steady
condition that develops is one of a broad, nearly {\it laminar} layer
separating the two counter-flowing fluids.  That difference is a
consequence of the magnetic field aligned with the original shear
layer.  For the stronger magnetic field case this result is an obvious
expectation, since magnetic tension stabilizes the instability before
any vortices can develop and the flow relaxes in a straightforward
manner to one that qualitatively resembles the initial conditions,
except that the shear layer is broadened until it is stable against
further perturbations. Because of the symmetries in our computations
the mean vector magnetic field is a constant throughout each
simulation, so the final magnetic field in this case is very little
different from that at the beginning. The total magnetic energy, for
example is the same as at the start to within 1\%. By the end of our
simulation there is a rough equipartition between magnetic and kinetic
energies, although both are small compared to the thermal energy
contained within the gas.

That the quasi-steady, relaxed flow state
should also be laminar for the weaker
magnetic field is perhaps not so immediately obvious, since the
initial magnetic energy is almost an order of magnitude less than the
kinetic energy and more than an order of magnitude less than the
thermal energy. It never becomes globally more than 5\% of the total
energy and never reaches global equipartition with the kinetic
energy. What happens is that as vortices do develop within the flow,
magnetic stresses are built up that have sufficient strength to locally
modify the dynamics.  At the same time magnetic field wrapped into the
vortices becomes unstable to tearing mode reconnection events that
isolate some magnetic flux and lead to large scale reorganization of
other magnetic structures.  The isolated magnetic flux is subsequently
annihilated. Together these developments produce a reorganized flow
and magnetic field in which the magnetic and flow fields are almost
perfectly aligned.  The magnetic field is also approximately as it was
at the beginning in this case, except for a pair of stronger flux structures,
or flux tubes, and fluctuations within them that appear to
correspond to Alfv\'en waves propagating in the same direction as the
fluid.  However, close examination reveals that the magnetic energy is
actually significantly greater than it was at the beginning. That
difference comes through the formation the
flux tubes. They formed out of an early magnetic reconnection
episode that magnetically isolated the shear layer from adjacent
plasma. Those flux tubes surround a sheet of hot, low density gas that
contains most of the excess entropy generated through the various
reorganization events leading to relaxation. Thus, through the action
of the magnetic field the flow has been reorganized into a thin, hot
entropy layer surrounded by magnetically enhanced structures within a
broad laminar shear layer.  In this case the shear layer is much
broader than for the stronger field case and actually substantially
broader than the width of the single, ``Cat's Eye'',
vortex that forms in the pure
gasdynamic case. The last distinction results from the fact that
magnetic reconnection following the initial generation of the
vortex lead to a significant impulsive transfer of both energy
and momentum away
from the initial shear layer.

Even though our ``weak field'' case had a magnetic field
initially dynamically fairly small, it was strong enough that before
the primary vortex could become fully developed, magnetic stresses
became {\it locally} important. It is reasonable, therefore, to ask
what differences should be expected when the initial field is
initially truly very weak.
Some discussions of MHD turbulence
in this very weak field regime (e.g., \cite{tingetal}) have argued
that magnetic fields will never become dynamically important; i.e, the
fields remain weak and passive.
But, we suspect that the answer depends on the effective
magnetic Reynolds of the flow, $R_m = U_oL/\eta$, comparing the
rate at which magnetic field is rolled into a vortex to the rate at
which it diffuses out of the vortex.
The simulations in the \cite{tingetal}
study, for example, had modest magnetic Reynolds numbers, $< 100$. On the other
hand, as pointed out, for example, by \cite{catvain91} and
\cite{cat94}, when the magnetic Reynolds number is big and there is a
large scale magnetic field, field line stretching can produce locally
strong fields. In astrophysical flows we may often anticipate very
large magnetic Reynolds numbers, so it is reasonable to expect in
cases where the large scale field starts out being quite small that it
might eventually play a role in determining the outcome (This point
has been made quite clearly with regard to MHD turbulent transport
(\cite{catvain91})).
As applied to
the present problem, this would mean that the flows might exist longer
as pseudo-gasdynamic, but that eventually field lines would be
stretched sufficiently that magnetic tension would become locally involved in
evolution of the vortex.
This process should also, of course, naturally create tearing
mode unstable magnetic topologies (e.g., \cite{biswel89}), whose growth
rates depend on the magnetic Reynolds number as well (e.g., \cite{mel86}).
That would provide a limiting process to accompany the eventual feedback
on dynamical stresses.
So, an outcome similar to what we have observed in the two cases studied here
is still reasonable for very weak fields
when the magnetic Reynolds number is large enough.
In practical problems, where flow conditions may be more complex or time
variable on large scales, these characteristics would have to be balanced
against time and length constraints imposed by those features, of course.

\acknowledgments
This work by AF, TWJ, and JBG was supported in part by the NSF through
grants AST-9100486 and AST-9318959, by NASA through grant NAGW-2548 and
by the University of Minnesota Supercomputer Institute.
The work by DR was supported in part by the Basic Science Research
Institute Program, Korean Ministry of Education 1994, Project No.
BSRI-94-5408.
We are grateful to Ibrahim Hallaj for very helpful assistance in
analyzing the results. We also appreciate some very constructive
comments on the manuscript by Akira Miura and Andrea Malagoli.

\clearpage

\begin{planotable}{lrrrrrrr}
\tablewidth{0pc}
\tablecaption{Calculation Initial Parameters}
\tablehead{
\colhead{Model}    &
\colhead{$U_o$}    &
\colhead{$B_o$}    &
\colhead{$a$}      &
\colhead{$M_s$}    &
\colhead{$M_a$}    &
\colhead{$\beta$}  &
\colhead{$\Gamma$} }
\startdata
${\rm strong~field}$ & $1$ & $0.4$ & $L/25$ & $1$ &
$2.5$ & $7.5$ & $0.053(U_o/2a)$ \nl
${\rm weak~field}$   & $1$ & $0.2$ & $L/25$ & $1$ &
$5$   & $30$  & $0.108(U_o/2a)$ \nl
\end{planotable}

\clearpage

\begin{center}
{\bf FIGURE CAPTIONS}
\end{center}
\begin{description}

\item[Fig.~1]
{The evolution of spatially averaged root mean square transverse
velocities normalized by their initial values.  Shown are the natural logarithm
of  the square root of the
average $u_y^2$
throughout the evolution of
the: (a) strong field case; (b) weak field case.  The solid, dotted, and
dashed-dotted lines correspond to the high ($512^2$), medium
($256^2$), and low ($128^2$) resolution simulations respectively.}

\item[Fig.~2]
{Gas density, (a), and magnetic field lines, (b), at 6 times
in the evolution of the high
resolution strong field simulation.  The gas density is shown as
greyscale images where dark correspond to low values.
To facilitate visualization of periodic structures across the boundaries
each image has been doubled. Coordinate values, measured from the lower
left are in units of $L$.
The times shown are 3.0, 6.0, 9.0, 11.9,
14.9 and 17.9 in units of $\tau= t/\Gamma^{-1}$.}

\item[Fig.~3]
{Gas density, (a), magnetic field lines, (b), and velocity field,
(c), at 6 times in the
evolution of the highest resolution weak field simulation.  Presentation
is similar to Fig. 2.
The velocity field is shown as vectors. The times shown are 2.4, 5.6,
8.0, 10.4, 14.4 and 20.0 in units of $\tau= t/\Gamma^{-1}$.}

\item[Fig.~4]
{Greyscale images (top to bottom) of the entropy, gas and
magnetic pressures at the end
of the weak field run (t = 20 $\tau$). Dark correspond to low
values.  To facilitate visualization on the periodic space each image has
been tripled.}

\item[Fig.~5]
{Fractional difference in magnetic energy between initial and final
states in simulations as a function of resolution.  Shown are
${{E_{bf} - E_{bo}} \over E_{bo}}$ where the subscripts ``f'' and ``o''
refer to the final and initial values of the magnetic energy.  Note
the units on the abscissa in the strong field case.}

\item[Fig.~6]
{The evolution of the spatially averaged squared magnetic field components for
the highest resolution weak field simulation.  Shown are the
values of $\langle (B_{k} - B_{ko})^2\rangle$, where $k=x,y$, are
indices. They are normalized to
the initial $x$ component squared, $(B_{xo})^2$.  The solid, and
dotted lines show the evolution of the $x$, and $y$ components
respectively.}

\item[Fig.~7]
{Evolution of energy components and $\beta$ parameter for the strong field
field case simulations.  Shown are spatially averaged values of the
thermal $E_t$, kinetic $E_t$, and magnetic energies $E_b$ (normalized
to the total energy $E_{Tot}$ = $E_t$ + $E_k$ + $E_b$).  In addition we
show the evolution of the minimum value on the computational grid of the
plasma parameter $\beta = {p_g \over p_b}$.  The solid, dotted, and
dashed-dotted lines correspond to the high ($512^2$), and medium
($256^2$) resolution simulations respectively.}

\item[Fig.~8]
{Same as Fig. 7 for the weak field case simulations. The dashed-dotted
lines corresponds to the low resolution simulation.}

\item[Fig.~9]
{Evolution of the secondary vortex.  Shown are the value of the surface
integral of vorticity and line integral of magnetic tension
in equation (3.6).  Note that the area integral was carried out
over the vortex and the line integral surrounded it.  The vorticity is
positive inside the area, so the signs indicate a reduction in the strength
of the vortex.}

\item[Fig.~10]
{Evolution of the cross helicity for
the high resolution weak field simulation.  Shown are the values of
$|\langle H\rangle |$ where $H = \int {\hat{\bmit u}} \cdot
{\hat {\bmit B}}dxdy$.  The dashed (dotted) line
corresponds to the top (bottom) half-planes.}

\end{description}

\end{document}